\input harvmac
\def \td {\tilde}
\def \const {{\rm const}}

\def\a{\alpha}

\def\g{\gamma}

\def\r{\rho}

\def\te{\theta}

\def\p{\phi}

\def \r {\rho}\def \g {\gamma}  
\def \ov {\over }
\def \P {\Phi}  \def \const {{\rm const}}
\def \q {{\tilde q}}
\def \ha {{1 \ov 2}}

\lref\KT{
I.~R.~Klebanov and A.~A.~Tseytlin,
``Gravity Duals of Supersymmetric $SU(N) \times SU(N+M)$ Gauge Theories,''
{\it Nucl. Phys.} {\bf B578} (2000) 123,
hep-th/0002159.
}

\lref\Buch{A. Buchel, ``Finite temperature resolution of the 
Klebanov--Tseytlin singularity,'' hep-th/0011146.
}

\lref\KS{I.~R.~Klebanov and M.~J.~Strassler, 
``Supergravity and a Confining Gauge Theory:  
Duality Cascades and $\chi$SB-Resolution of Naked Singularities,''
{\it JHEP} {\bf 0008} (2000) 052,
hep-th/0007191.
}

\lref\PT{L. A.  Pando Zayas and A. A. Tseytlin,
``3-branes on spaces with $R \times  S^2 \times  S^3$ topology,''
hep-th/0101043.
}

\lref\PTs{L.~A.~Pando Zayas and A.~A.~Tseytlin,
``3-branes on resolved conifold,''
{\it JHEP} {\bf 0011} (2000) 028,
hep-th/0010088.
}

\lref\SK{A.~Kehagias and K.~Sfetsos,
``On running couplings in gauge theories from type-IIB supergravity,''
{\it Phys. Lett.} {\bf 454B} (1999) 270,
hep-th/9902125;
``On asymptotic freedom and confinement from type-IIB supergravity,''
{\it Phys. Lett.} {\bf 456B} (1999) 22,
hep-th/9903109;
S.~Nojiri and S.~D.~Odintsov,
``Two-boundaries AdS/CFT correspondence in dilatonic gravity,''
{\it Phys. Lett.} {\bf 449B} (1999) 39,
hep-th/9812017.
}

\lref\Gub{S.~S.~Gubser,
``Dilaton-driven confinement,''
hep-th/9902155.
} 

\lref\gim{G.~W.~Gibbons and K.~Maeda,
``Black Holes And Membranes In Higher Dimensional Theories With Dilaton Fields,''
{\it Nucl. Phys.} {\bf B298} (1988) 741.
}

\lref\dup{  M.~J.~Duff, H.~Lu and C.~N.~Pope,
``The black branes of M-theory,''
{\it Phys. Lett.} {\bf 382B} (1996) 73,
hep-th/9604052.
 } 

\lref\call{ C.~G.~Callan, S.~S.~Gubser, I.~R.~Klebanov and A.~A.~Tseytlin,
``Absorption of fixed scalars and the D-brane approach to black holes,''
{\it Nucl. Phys.} {\bf B489} (1997) 65,
hep-th/9610172.
}

\lref \HS {G.~T.~Horowitz and A.~Strominger,
``Black strings and P-branes,''
{\it Nucl. Phys.} {\bf B360} (1991) 197.
}

\lref\Cvetic{
M.~Cvetic, H.~Lu and C.~N.~Pope,
``Brane resolution through transgression,''
hep-th/0011023.
M.~Cvetic, G.~W.~Gibbons, H.~Lu and C.~N.~Pope,
``Ricci-flat metrics, harmonic forms and brane resolutions,''
hep-th/0012011.
}

\lref \HK {C.~P.~Herzog and I.~R.~Klebanov,
``Gravity duals of fractional branes in various dimensions,''
hep-th/0101020.
}

\lref\GKP{S.~Gubser, I.~R.~Klebanov and A.~Peet,
``Entropy and temperature of black 3-branes,''
{\it Phys.~Rev.~D} {\bf 54} (1996) 3915, hep-th/9602135.
}

\lref\Witt{E.~Witten,
``Anti-de Sitter space, thermal phase transition, and confinement in gauge
theories,''
{\it Adv.~Theor.~Math.~Phys.} {\bf 2} (1998) 505, hep-th/9803131.
}

\Title{\vbox
{\baselineskip 10pt
{\hbox{PUPT-1977}\hbox{MCTP-01-07}  \hbox{OHSTPY-HEP-T-01-002}\hbox{NSF-ITP-01-10}
\hbox{hep-th/0102105}
}}}
{\vbox{\vskip -30 true pt\centerline {
Non-Extremal Gravity Duals for }
\medskip
\centerline { Fractional D3-Branes on the Conifold  }
\medskip
\vskip4pt }}
\vskip -20 true pt
\centerline{A. Buchel$^{1}$,
C. P. Herzog$^{2}$, I. R.~Klebanov$^{1,2}$, 
L. Pando Zayas$^{3}$ and A. A. Tseytlin$^{4,5}$\footnote{$^*$} {Also at Lebedev 
Physics Institute, Moscow. }
 }
\smallskip\smallskip
\centerline{$^{1}$ \it Institute for Theoretical Physics,
University of California, Santa Barbara, California 93106-4030}
\centerline{$^{2}$ \it Joseph Henry
Laboratories, Princeton University, Princeton, New Jersey 08544}
\centerline{$^{3}$ \it  Michigan Center for Theoretical Physics, 
The University of Michigan, Ann Arbor, MI 48109}
\centerline{$^{4}$ \it  Department of Physics, 
The Ohio State University,  
Columbus, OH 43210 }
\centerline{$^{5}$ \it  Blackett Laboratory, 
Imperial College,   London SW7 2BZ, U.K. }

\bigskip\bigskip
\centerline {\bf Abstract}
\baselineskip12pt
\noindent
\medskip
The world volume theory on
$N$ regular and $M$ fractional D3-branes at the conifold
singularity
is a non-conformal ${\cal N}=1$ supersymmetric $SU(N+M)\times
SU(N)$
gauge theory. In previous work the extremal Type IIB supergravity dual of this
theory at zero temperature was constructed.
Regularity of the solution requires a deformation of the conifold:
this is a reflection of the chiral symmetry breaking.
 To study the non-zero temperature
generalizations non-extremal solutions have to be considered, and
in the high temperature phase the chiral symmetry is expected to be restored.
Such a solution is expected to have a regular Schwarzschild horizon.
We construct an ansatz necessary to study such non-extremal solutions
and show that the simplest possible solution has a singular horizon.
We derive the system of second order equations in the radial variable 
whose solutions may have regular horizons.
\bigskip

\Date{02/01}

\noblackbox 
\baselineskip 15pt plus 2pt minus 2pt

\newsec{Introduction}

In this paper we study non-extremal generalizations of
the KT solution \KT, which describes regular and fractional D3-branes
at the apex of the conifold.
The extremal KT solution is crucially dependent on the presence of 
Chern-Simons terms in type IIB supergravity.
These terms cause the RR 5-from flux to vary radially.
Indeed, while in regular D3-brane
solutions $dF_5=0$, here the 3-form field strengths 
are turned on in such a way that the right-hand side of the
equation
\eqn\CS{
dF_5 = H_3\wedge F_3\ 
}
does not vanish. In fact, the 5-form flux increases without bound for large
$r$. In \KS\ this behavior was attributed to a cascade of Seiberg
dualities in the dual ${\cal N}=1$ supersymmetric $SU(N)\times SU(N+M)$
gauge theory.

Further examples of supergravity backgrounds with varying flux were
constructed in \refs{\KS,\PTs,\Cvetic,\HK}. 
In particular, it is important to understand the resolution
of the naked singularity present in the KT solution. The proposal
of \KS\ is that the conifold becomes deformed. 
As a result of this deformation the extremal
KS solution is perfectly non-singular and
without a horizon in the IR, while it asymptotically approaches
the KT solution in the UV (for large $\r$).
The mechanism that removes the naked singularity is related to the
breaking of the chiral symmetry in the dual $SU(N)\times SU(N+M)$
gauge theory. The $Z_{2M}$ chiral symmetry, which may be approximated by
$U(1)$ for large $M$, is broken to $Z_2$ by the deformation of
the conifold \KS.

In \Buch\ a different mechanism for resolving this naked singularity
was proposed. It was suggested that a non-extremal generalization
of the KT solution may have a regular Schwarzschild horizon
``cloaking'' the naked singularity. The dual field theory
interpretation of this would be 
the restoration of chiral symmetry at a finite temperature
$T_c$ \Buch. One expects that
turning on finite temperature in the field theory, which translates
into non-extremality on the supergravity side \refs{\GKP,\Witt},  
leads to restoration
of the chiral symmetry above some critical temperature $T_c$.\foot{
Strictly speaking, we cannot {\it a priori} rule out the possibility that
$T_c=\infty$. In that case, a regular black hole in KT geometry does not
exist. While this possibility seems strange, the only way to decide
the issue is to find the actual  black hole solution where 
a regular Schwarzschild horizon  does shield the singularity.}
The 
symmetry restoration is 
part of the deconfinement transition at $T_c$ 
for this particular ${\cal N}=1$
gauge theory. 
The proposal of \Buch\ is that the description of the 
phase with restored symmetry involves a regular Schwarzschild horizon
appearing in the asymptotically KT geometry. This proposal
is analogous to the fact that the ${\cal N}=4$ SYM theory,
which is not confining, is described at a finite temperature by
a black hole in $AdS_5$ \GKP,\Witt. 
The difference is that in our case the $T=0$ theory exhibits 
confinement and chiral symmetry breaking \KS. So, the regular Schwarzschild
horizon should appear only for some finite Hawking temperature.
This would be a rather unusual and, to our knowledge, unstudied phenomenon
from the supergravity point of view.

One implication is that, 
at temperatures below $T_c$, 
there should be non-extremal generalizations
of the KS solution which are free of horizons, just like the
extremal solution. The absence of a horizon is a manifestation
of confinement, as evidenced
by the resulting area law
for Wilson loops or, alternatively, 
by counting of degrees of freedom.
Once the horizon appears, the Bekenstein-Hawking entropy associated with
it typically scales as $N^2$ where $N$ is the relevant number of colors.
This factor indicates that the color degrees of freedom are not confined.
If there is no horizon, then the entropy could only appear through
string loop effects which would make it of order $N^0$, in
agreement with color confinement. 
It is interesting, therefore, to study this theory
as a function of the temperature, and to identify the phase transition
where the chiral symmetry is restored.

In order to address these questions from
a dual supergravity point of view, we need to study non-extremal
generalizations of the KT and KS backgrounds. 
This problem was first addressed for the KT background
in \Buch\ 
partly with numerical methods. We rederive this solution analytically
and show that the identification of the horizon as $r=r_*$ in eq. (2.60)
of \Buch\ is incorrect because $\Delta_1(r_*)\neq 0$. 
The vanishing of $\Delta_2(r_*)$ is an artefact of the coordinate choice.
We find a good radial coordinate $u$ and show that the solution
derived in \Buch\
has a singular horizon at $u=\infty$ which corresponds to $r=\infty$
on another branch of the solution. This type of singular horizon 
is deemed unacceptable in studies of black hole metrics.

Thus,  
the solution found in \Buch\ does not have a regular
Schwarzschild horizon shielding the naked singularity of the extremal KT
metric for sufficiently high Hawking temperature. 
Instead, the horizon (defined as the locus where 
$G_{00}=0$) and the singularity are
coincident, independent of the choice of the non-extremality parameter.
We also show that, in the limit where we remove 
the fractional D3-branes (wrapped D5-branes), and leave
only the regular D3-branes, the solution of \Buch\ does not reduce to the
standard non-extremal 3-brane metric. 
Instead, it reduces to a non-standard non-extremal version of a D3-brane solution whose 
metric has  a singular horizon.

Nevertheless, the scenario for 
chiral symmetry restoration proposed in \Buch\ is very appealing.
This motivates us to introduce a more general $U(1)$ symmetric
ansatz and to begin search for solutions
that are asymptotically KT but possess regular Schwarzschild horizons.

\newsec{Non-Extremal Generalization of the KT Ansatz}

We start with an ansatz for the non-extremal KT
background. Just as in \KT\ we impose the requirement that
the background has a $U(1)$ symmetry associated with the $U(1)$
fiber of $T^{1,1}$. Our ansatz will be more general than that of \Buch.
 It turns out that in order to look for
solutions which reduce to the standard non-extremal D3-brane
in the limit of zero fractional brane charge $P$, one 
should not 
 impose self-duality on the 3-forms away from extremality.
This in turn implies 
that one 
is to adopt a more general  ansatz  for the metric 
 than in \Buch\ and also 
allow for a non-constant dilaton.

A general ansatz for a 10-d  Einstein-frame metric
 consistent with the $U(1)$ symmetry of $\psi$-rotations
and the interchange of the two $S^2$'s involves 
4 functions $x,y,z,w$  of a radial coordinate $u$
\eqn\mett{
ds^2_{10E} =  e^{2z} ( e^{-6x} dX_0^2 + e^{2x} dX_i dX_i)
+ e^{-2z}  ds^2_6 \ ,   }
where
\eqn\mott{ds^2_6 =  e^{10y} du^2 + e^{2y} (dM_5)^2  \ , } 
\eqn\mmm{
(dM_5)^2 =  e^{ -8w}  e_{\psi}^2 +  e^{ 2w}
\big(e_{\theta_1}^2+e_{\phi_1}^2 +
e_{\theta_2}^2+e_{\phi_2}^2\big) \equiv
 e^{ 2w} ds_5^2    \ , }
and 
$$
 e_{\psi} =  {1\ov 3} (d\psi +  \cos \theta_1 d\phi_1  +  \cos \theta_2 d\phi_2)  \  , 
 \quad  e_{\theta_i}={1\ov \sqrt 6} d\theta_i\ ,  \quad  e_{\phi_i}=
{1\ov \sqrt 6} \sin\theta_id\phi_i \ .
$$
Here $X_0$ is the euclidean time and $X_i$ are the 3 longitudinal 3-brane directions.

This metric  can  be brought into a more familiar D3-brane form
\eqn\fop{
ds^2_{10E} =  h^{-1/2}(\r)  [ g(\r)  dX_0^2 + dX_i dX_i]
+  h^{1/2}(\r)   [  {\rm g}^{-1} (\r) d\r^2 
+ \r^2  ds^2_5] \ , } 
with the redefinitions 
\eqn\iop{
 h=  e^{-4z- 4x} \ , \ \ \ \ \     \r  = e^{y + x + w  }\ ,
 \ \ \
\ \ \     g=  e^{-8x}\  ,\ \ \ \ \ \ \ e^{10y + 2x} du^2  = {\rm g}^{-1} (\r) d\r^2 \ .  }
When $w=0$ and $e^{4y}=\rho^4= { 1 \ov 4u} $,
the transverse 6-d space is the standard conifold
with $M_5= T^{1,1}$.  
Small  $u$  thus  corresponds to large distances
(where we shall assume that $h,g,v\to 1,$ as $\r \to \infty$)
 and vice versa.

The function $w$ squashes
the $U(1)$ fiber of $T^{1,1}$ relative to the 2-spheres; it does not
violate the $U(1)$ symmetry.
A  Ricci-flat  6-d space  with non-trivial  $w$  is 
the generalized conifold of \PT\
\eqn\genn{
ds_6^2={ \kappa^{-1}(\r)}d\r^2+  \r^2 \big[ 
\kappa (\r) 
e_{\psi}^2 +  e_{\theta_1}^2+e_{\phi_1}^2+
e_{\theta_2}^2+e_{\phi_2}^2\big] ,
}
\eqn\kaa{
\kappa(\r)=  e^{-10w} = 1-{\r_*^6 \over \r^6} \ , \ \ \ \  \ \ \
\r = e^{ y +  w } , \ \ \ \     \r_*   \leq \r < \infty  \ . }
This  space has regular curvature, and 
a bolt singularity at $\r=\r_*$ 
is removed by  $Z_2$ identification of the angle $\psi$.
The limit of the standard conifold is $\r_* \to 0$ or $y_* \to  -\infty$
which corresponds   to $w=0$.

The extremal D3-brane on the conifold and the more general fractional D3-brane   
KT solution have  $x=w=0$ 
(for their $w\not=0$ analogs  in the case  when the 6-d space is 
the generalized conifold see \PT).
Adding a non-constant $x(u)$  drives the non-extremality. For example, 
the non-extremal version of a D3-brane on a standard ($w=0$) 
conifold solution 
has $x=au, \ \  e^{-8x} = g= 1 - { 2a\ov \r^4}, \ e^{4z+4x} = h=1 + 
{\td q \ov \r^4}, \  \r=e^{y+x}.$
Our aim will be to understand how switching on 
the non-extremality ($x=au$)  changes the extremal KT solution.

Our ansatz for the $p$-form  fields is dictated by 
symmetries  and thus  is  exactly the same 
as in the extremal KT case \KT:\foot{
The function $T$ in \KT\ is  related to $f$ by 
$f= { 1 \ov \sqrt 2 } T$, and we make similar rescaling of $P$:
$P = { 1 \ov \sqrt 2 }  P_{KT} $.
}
\eqn\har{
F_3 = \   P
e_\psi \wedge
( e_{\theta_1} \wedge e_{\phi_1} - 
e_{\theta_2} \wedge e_{\phi_2})\ ,
} 
\eqn\ansa{
B_2  = \    f(u) 
( e_{\theta_1} \wedge e_{\phi_1} - 
e_{\theta_2} \wedge e_{\phi_2})
 \ , 
}
\eqn\fiff{
F_5= {\cal F}+*
{\cal F}\
, \quad  \ \ \ \ {\cal F} = 
K(u) 
e_{\psi}\wedge e_{\te_1} \wedge
e_{\p_1} \wedge
e_{\te_2}\wedge e_{\p_2}\ . }
As in \KT, the  Bianchi identity for the
5-form, \ $
d*F_5=dF_5=H_{3}\wedge F_3$,  implies
\eqn\kee{
K (u)  = Q + 2 P f (u) \ .}

In what follows, we shall derive the 
corresponding  system of type IIB supergravity equations of motion 
 describing 
radial  evolution of the six unknown functions of $u$ --
$x,y,z,w,f$ and  $\P$.
We shall then discuss its 
solutions generalizing the work in \KT\ to the non-extremal case.

The simplest special fixed-point solution of our system 
turns out to have $w=0$, 
i.e.~the case when $T^{1,1}$ is not squashed.  
If the fractional branes are present, then the only $U(1)$
symmetric solutions with
$w=\Phi=0$ are 
the KT solution \KT\ 
and its non-BPS generalization considered in \Buch.
We will  discuss  the latter solution 
in some detail in sec.~4, 
finding its  {\it explicit} analytic form and clarifying its geometry. 
Its horizon turns out to be singular. A related problem is 
that it does not reduce to the regular non-extremal D3-brane
solution in the limit of no fractional brane charge, $P\to 0$.

Demanding that the non-extremal solution have the correct $P=0$ limit 
leads to the necessity 
of relaxing the 
condition  $f'= -P e^{\P+ 4y }$, i.e.~that the 3-forms
are self dual.
If the 3-forms are not self-dual, then
$\Phi$ and $w$  can  no longer 
be held constant. In the next section we derive the resulting system
of second-order differential equations.

\newsec{Basic  Equations}

\subsec{\bf Effective  1-d action for radial evolution }
As in \KT, the most efficient way to derive the
  system of type IIB supergravity equations of motion 
is to  follow \refs{\gim,\call} and to 
start with the 1-d effective action 
for the radial evolution.

For the metric \mett\ $ \sqrt G={1 \ov 108}  e^{10y -2 z}$
(up to angle-dependent factors), and  
computing the scalar curvature 
we find 
\eqn\gre{
\int d^{10} x \sqrt G  R\ \  \to \ \  \ \  
 { 1 \ov 27}\int du \big[ 5 y'^2 - 3 x'^2 - 2 z'^2  - 5 w'^2
+  e^{8y} ( 6 e^{-2w} - e^{-12 w} )  \big] \ . 
}
Note that  $w=0$  is a consistent fixed point of the equations of motion.  
Replacing $M_5$  in \mmm\
by the standard  $T^{1,1}$  or 
by $S^5$ produces 
exactly the same 1-d gravitational  lagrangian.  
That means, in particular,  that {\it for $w=0$}
 the regular D3-brane  solution  and its 
{\it  non-extremal } version 
will not change if we replace the flat transverse space 
${\bf R}^6$ with the conifold.

It is easy to see that the matter part $L_m$ of the effective lagrangian
is  essentially the same as it was in the extremal case \KT\ 
(apart from $w$-dependent factors): $L_m$ does not depend on the
non-extremality function $x$. Thus,  following  \KT, 
$$\int d^{10} x \sqrt G \ [ - { 1 \ov 2} ( \del\P)^2 +...] \ \  \to \ \  $$
\eqn\maat{
  - { 1 \ov 27}\int du \    { 1 \ov 8} \bigg[ 
 \Phi'^2 + 2 e^{-\Phi + 4z -4y - 4 w} f'^2 
 + 2  e^{\Phi + 4z + 4y + 4w} P^2 + 
  e^{8z} (Q + 2 P f)^2   \bigg]\ . 
}
{}{}From \gre\ and \maat\  we 
get the following 1-d effective lagrangian
(ignoring an irrelevant overall  numerical factor)
$$
L=   T  - V \  , $$ 
\eqn\tki{ T =  5 y'^2  -  3 x'^2 - 2 z'^2  - 5 w'^2 
- { 1 \ov 8} \P'^2
- { 1 \ov 4}  e^{-\P +  4z -4y - 4 w } f'^2  
\ , 
 } \eqn\lagg{  
 V = -    e^{8y} ( 6 e^{-2w} - e^{-12 w} )
 +  { 1 \ov 4} e^{\P+  4z + 4y + 4 w } P^2 + 
 { 1 \ov 8}  e^{8z} (Q + 2 P f)^2 \ , 
}
supplemented with 
the ``zero-energy"  constraint $T+V=0$. 

Since the non-extremality function $x$
does not appear in $\sqrt G$, $G_{uu}$, or 
the angular part of the metric, 
it is absent in $L_m$ and thus is a ``modulus'' 
-- it has no potential (cf. \gre). Thus 
its dependence on $u$ is simply linear
\eqn\xex{
x''=0 \ , \ \ \ \  x= a u \ , \ \ \ \     a=\const > 0 . }
This is just what is expected of an extra kinetic 
energy (i.e. the $ x'^2= a^2$ term appearing in the 
  ``zero-energy"  constraint)
which spoils the BPS nature of the KT solution for 
a non-constant $x$.

\subsec{\bf The Superpotential and the extremal KT solution}

The crucial observation made  in \KT\ is that 
 the lagrangian \tki, \lagg\ has a  remarkable  special structure -- 
it admits a superpotential. 
Indeed, it can be 
 be represented in the following  way
$$  L  =  - 3 x'^2  - { 1 \ov 8} \P'^2  $$  $$  
+ \ 5 [y' + { 1 \ov 5} e^{4y}(3 e^{4 w} + 2   e^{-6w} )]^2 
 -\ 5  [w' - { 3 \ov 5} e^{4y} (e^{4w} -  e^{-6 w})]^2 $$
\eqn\laa{ -\ 
2 [z' + { 1 \ov 4} e^{4z} (Q + 2 P f)]^2
- { 1 \ov 4}  e^{-\P +  4z -4y-4w} (f'  + P e^{\P + 4y + 4 w}  )^2
 } 
$$ 
 -\   2 
\bigg[ { 1 \ov 4} e^{4y} (3 e^{4 w}  +  2 e^{-6w}  )
 - { 1 \ov 8} e^{4z}  (Q + 2 P f)\bigg]' \ , 
$$ 
where the last term is a total derivative  and  may be dropped.
Let us recall that, if,  in general,    $V(\p)$ 
can be expressed in terms of  a function $W(\p)$ as 
$V =  - g^{ij}  \del_i W  \del_j W   , $
where $g_{ij}(\p)$ is  the kinetic term
metric, 
 then 
\eqn\loi{L= T-V= g_{ij}(\p)  \p'^i \p'^j - V(\p) 
= g_{ij} (  \p'^i + g^{ik} \del_k W ) ( \p'^j + g^{jl} \del_l W) 
  - 2 W'
\ . } 
As a result, there exists a 
 special BPS  solution of the  corresponding 
2-nd order equations, 
satisfying     
\eqn\syy{  \p'^i + g^{ik} \del_k W  = 0 \ ,  }
and thus also the zero-energy constraint.
As follows from \laa, in 
 the present case  \KT
\eqn\supp{
W=   { 1 \ov 4} e^{4y} (3 e^{4 w}  +  2 e^{-6w}  )
 - { 1 \ov 8} e^{4z}  (Q + 2 P f)
\ . }
Note that $W$ does not depend on $\P$. 
The corresponding system \syy\  of 1-st order equations is
then  \KT:
\eqn\bops{
x'=0\ , \ \ \ \ \ \  \P'=0 \ ,}
\eqn\bps{
y' + { 1 \ov 5} e^{4y}(3 e^{4 w} + 2   e^{-6w} )       =0 \ , \ \ \ \
\ \ \ \ \
w' - { 3 \ov 5} e^{4y} (e^{4w} -  e^{-6 w})       =0 \ ,  }
\eqn\zf{ f'  + P e^{\P + 4y+4w}  =0 \ , \ \ \ \ \ \ \ \ \ 
z' + { 1 \ov 4} e^{4z} (Q + 2 P f)=0\ . }
The equation for $f$ in \zf\ 
implies self-duality of the complex
3-form field,  i.e. $H_3 = e^\Phi \star F_3$.

Choosing the special solution $w=0$  of the $w$-equation in 
\bps\ 
 we then  find  the  KT solution   \KT\ 
 $$  x=0\ , \ \ \ \ \ \ \ w=0\ , \ \ \ \ \P=0 \ , $$ 
\eqn\kot{  e^{-4y} = 4u \ , \ \  \ \ \ \ \ \ 
\ f= f_0  - {P\ov 4} \ln u \ ,   } 
  $$ e^{-4z} = 1 + K_0 u  
 -  {P^2\ov 2}  u ( \ln u -1) \ ,  \ \ \ \ \ \ \   
 K_0 = Q + 2 P f_0 \  , $$ 
i.e. 
\eqn\ktt{ e^{-4z} = h =  
 1 +   (Q + 2 P f_0 + {P^2\ov 2} ) u  
 -  {P^2\ov 2}  u \ln u \ .      }
In terms of the radial  variable $\r$ used 
 in \KT, 
$$  e^{4y}= \rho^4 = { 1 \ov 4u}  \ , \ \ \ \ \ \ \ 
f=  f_1  +  {P} \ln \r  \ ,   $$ 
\eqn\ktr{
 h= 1 +   (Q + 2 P f_1  + {P^2\ov 2}  
) { 1 \ov 4\r^4}  
 +  {P^2\ov 2\r^4 }  \ln \r 
\ , } 
where $f_1= f_0 + { P \ov 2} \ln 2$. 

A more general extremal solution of \bops--\zf\ 
with non-zero $w$ 
leads to fractional D3-branes on the generalized conifold 
\genn\ solution of \PT.\foot{ For its  explicit form 
in terms of the $\r$ coordinate, 
see eqs. (4.20), (4.21) in \PT. While the generalized conifold 
has, in contrast to the standard conifold, 
regular curvature, the back reaction of D3-branes 
makes the metric singular:
pure  D3-branes  on the generalized conifold  have a horizon 
coinciding with the singularity. The fractional D3-brane solution is similar to the KT one:
it has a naked singularity behind the $K(u)=0$ locus.}

\subsec{\bf The full system of 2-nd order equations}
We would  like  to generalize the solution 
\ktt\ to the non-BPS case when $x'\not=0$, i.e.
the non-extremality parameter 
$a$ in \xex\ is non-zero.
In order to do that  we need to  start with 
 the original  2-nd order system  following from
\tki, \lagg\ or  \laa, i.e.~the free equation for $x$ \xex\ 
 plus  a coupled system for $y,w,z,f$ and $\P$ 
\eqn\yyy{ 10y'' - 8 e^{8y} (6 e^{-2w} - e^{-12 w})   + \P''
=0 \ , 
}
\eqn\yuw{
10w'' - 12 e^{8y} ( e^{-2w} - e^{-12 w})   - \P''
=0 \ , } 
\eqn\ppp{
\P''    + e^{-\P + 4z - 4y-4w} (f'^2 -  e^{2 \P + 8 y+8w} P^2)=0 \ , }
\eqn\zzy{
4z'' -  (Q+ 2 P f)^2  e^{8z}
 - e^{-\P + 4z - 4y-4w} ( f'^2 +  e^{2 \P + 8 y+8w} P^2) =0\ , 
}
\eqn\fef{
(e^{-\P + 4z - 4y-4w} f')' - P (Q+ 2 P f) e^{8z} =0 \ . 
}
The integration constants are 
subject  to the  zero-energy constraint $T+V=0$, i.e.
$$5  y'^2   - 2 z'^2  - 5 w'^2 - { 1 \ov 8} \P'^2 
- { 1 \ov 4}  e^{-\P +  4z -4y - 4 w } f'^2 $$ 
\eqn\coln{ 
 -   \  e^{8y} ( 6 e^{-2w} - e^{-12 w} )
 +  { 1 \ov 4} e^{\P+  4z + 4y + 4 w } P^2 + 
 { 1 \ov 8}  e^{8z} (Q + 2 P f)^2   = 3 a^2  \ . 
}
This system has special properties  reflecting 
the structure of the lagrangian \laa.
In particular, there is a  subclass of  simple 
solutions for which 
the 1-st order equations for $f$ and $z$ \zf\
are  still satisfied, while \bops\ and \bps\
are replaced by their  2-nd order counterparts.
Indeed, 
it is easy to see that if we set 
$f'^2 -  P^2 e^{2 \P + 8 y+8w}=0$, 
i.e.~$f'= -  P e^{ \P + 4 y+4w} $, then \fef\ 
implies that $z$ should be subject  to the first-order equation in \zf. 
In this case the 3-forms are self-dual.
Then it is consistent, in particular, 
 to keep $w=0$ so that $T^{1,1}$ is not squashed.
We will discuss this class of solutions  first
in the next section.

In general,  if we  relax  the 1-st order  conditions \zf\ 
on $f,z$   -- and we will 
be forced to relax them  in  order to have a regular horizon --
the dilaton will  run according to \ppp, driven by 
the non-extremality $x'=a>0$.
From \yuw, if we relax the first order constraint on $f$, 
the function $w$ will also be forced to run.  Hence, we need
the more general metric \mett\ in order to have a regular horizon.
These nontrivial dilaton and $w$ dynamics constitute
a  novel phenomenon  specific to the
non-extremal 
{\it fractional} D3-brane  case, $a>0$ and $P>0$.

\newsec{A Singular Non-BPS Generalization of the KT Solution}
Assuming  that $f$ and $z$  are subject to \zf, i.e. 
\eqn\zfe{ f'  + P e^{\P + 4y+4w}  =0 \ , \ \ \ \ \ \ 
z' + { 1 \ov 4} e^{4z} (Q + 2 P f)=0\ ,  }
so that \zzy\ and \fef\ are  satisfied automatically, 
 the remaining 
equations  \yyy, \yuw, \ppp\ and \coln\ become
\eqn\wnb{
5y'' - 4 e^{8y}  (6 e^{-2w} - e^{-12 w})     =0 \  ,} 
\eqn\wwnb{
 5w'' - 6 e^{8y} ( e^{-2w} - e^{-12 w})  =0 \ ,     }
\eqn\pop{ \P''=0 \ ,  }
and 
\eqn\tko{ 5 y'^2     - 5 w'^2  - {1 \ov 8} \Phi'^2 
 -    e^{8y} ( 6 e^{-2w} - e^{-12 w} ) = 3 a^2  
\ . }
Notice that the matter and gravity parts are now  decoupled:
eqs.~\wnb--\tko\  
 would  be obtained   just by looking for  Ricci-flat  
uncharged   black 3-brane   solutions with a metric 
in the class \mett. 
Finding first  the functions $y$ and $w$, one is
then to plug them back into \zfe\  to determine 
the functions $f$ and $z$.

The  BPS solution  of   \wnb--\tko\  
corresponds to $a=0$
  and $y,w$ subject to \bps,  leading to 
the generalized conifold space
\genn, \kaa.

We begin an analysis of the non-BPS solutions 
by discussing the  special case when $w=\Phi=0$.   
In this case, the $T^{1,1}$ 
is not squashed, and the eqs.~\wnb--\tko\ simplify substantially.  
When $P>0$ we will see that we
recover the non-BPS generalization of the KT solution  
first discussed in \Buch. 
 In section ~5.2  we will see that 
in the $P=0$ limit this solution does not become the standard (regular) non-extremal D3-brane solution.
Similar singular non-BPS solutions with non-constant dilaton
are  constructed in Appendix A.

Integrating \wnb\  and using \tko\ 
we find
\eqn\nnn{
y' = -\sqrt{ b^2 + e^{8y}}  \ ,\ \ \ \ \   
 \ \ \  b  = \sqrt{3\ov 5}  a  \ .  }
Assuming the required long-distance ($u\to 0$) 
asymptotic  conditions we then have
\eqn\wesq{
e^{4y} = { b \ov \sinh 4bu} \ .}
Integrating \zfe\ shows that 
 \eqn\fff{
f= f_*  -  {P\ov 4} \ln \tanh 2bu \ ,   
}
i.e. $f$ approaches a {\it constant} at large $u$
and has the KT behavior \kot\  at small $u$.

The large $u$ (short distance) 
behaviour is an improvement 
compared to the extremal KT case:
since $f$  stays positive (does not change sign) 
so does $K$ which is a derivative of $e^{-4z}$
(cf. \kee,\zfe).
That means that $e^{-4z}$ will keep growing at small distances, 
and does not go to zero.
The total function $h= e^{-4z-4x}$ 
in \fop\ will vanish only at $u=\infty$
due to  the vanishing of $e^{-4x}$, 
i.e. here the  {\it singularity  coincides with 
the horizon}. 

 This is a  rather peculiar situation, 
different from the  naked singularity of the KT solution.
The non-vanishing of the 
non-extremality  parameter $a \sim b$ is 
obviously crucial for this
difference.\foot{Other known cases of solutions
where a horizon coincides with a curvature singularity
include, for example, 
   dilatonic  BPS Dp-branes \HS\   as  well as  D3-branes 
 on generalized conifolds \PT.}

Explicitly,  from the equation for $z$ in \zfe\ 
$
e^{-4z} =   \int du\  [ Q + 2 P f(u) ]$, 
we find  that 
$z$ can be expressed in terms of polylogarithms
\eqn\poll{
e^{-4z} = C  + K_* u + { P^2 \ov 8 b} \bigg({\rm Li}_2(-e^{-4bu})
- {\rm Li}_2(e^{-4bu}) \bigg) \ , }
where  
$ K_* = Q + 2 P f_* \  $.\foot{Note that 
 $e^{-4z (0)} = C -  { \pi^2 P^2 \ov 32 b}$.
The constant $C$ may be 
adjusted so that
the solution has or does not have an 
 asymptotically flat region; the latter possibility corresponds to
 $h(0)=0$.} 

The exact analytical solution presented here realises the non-BPS 
KT background discussed in \Buch\ . In particular, the equation 
(2.53) in \Buch\ for the warp factor $\triangle_1(\tau)\equiv \sqrt{f(x)}$   
 that determines the position of the event horizon 
is solved with the identification\foot{Here $f$  and $x$ 
refer to the notation   in \Buch.}
\eqn\mapu{
x\equiv x(u)={4a\over 3b} e^{-4 a u}\ \sinh 4bu \ , }
\eqn\mapf{
f(x)\equiv f(x(u))=e^{-8 a u}\ , }
provided  we set $a={243\ov 4}  A$. 
In \Buch, $f(x_\star)$ was found 
numerically to vanish for $x_\star\ne 0$. Given 
\mapu,\ \mapf\  we see that this statement 
is incorrect. This numerical error led to the 
wrong conclusion of the nonsingular horizon of the black 
hole solution in the KT geometry proposed in \Buch\ .

Near $u=0$ we get from \poll\ the same expression as 
in the extremal case ($a=0,\ b=0$), i.e.~the 
KT behaviour \ktt\ 
where $e^{-4z}$ is 1 at $u=0$ and grows as $u$ increases.
For  large $u$:
 \eqn\seol{  y= - bu + { 1 \ov 4} \log 2b + { 1 \ov 4} e^{-8b u} + ...\ , \ \ \ \ \
f= f_*   +  {P\ov 2} e^{- 4bu} + ... \ , }
\eqn\zezf{
e^{-4z} = C +  K_*  u - { P^2 \ov 4b}   e^{- 4bu}  +  ...  \ . }
Thus  $e^{-4z}$ always grows never reaching zero.
If we define the horizon as the locus where $G_{00}=\exp(2z-6x)$ vanishes,
then from these large $u$ asymptotics, it is clear that we have a horizon
as $u \to \infty$.  Moreover, from \mett\ and the differential equations
\zfe, \wnb\ and \tko, we find that the Ricci scalar in this space-time
is $R = P^2 e^{6z - 6y}$. From the large $u$ asymptotics, 
it is clear that $R$ is
singular in the limit
$u\to \infty$.

We conclude that the non-BPS  solution
\wesq--\poll\  does not  have a horizon shielding 
the naked singularity  of the extremal KT solution;
instead, the introduction of non-extremality   here 
creates a horizon at $u=\infty$ and  shifts the singularity 
of the extremal KT background from
a finite value of $u$ to the same point $u=\infty$.

In fact, if we take the $P\to 0$ limit of \wesq--\poll,
thus removing the fractional branes,
we still have a singular horizon at $u=\infty$: even though
the Ricci scalar vanishes, components of the curvature tensor blow up.
The lesson is that
the presence of the fractional charge does not influence the 
singularity significantly.
We will show in the next section that this $P=0$ limit
corresponds {\it not} to the ordinary non-extremal D3-brane
solution but
to a special non-BPS D3-brane solution.

\newsec{
General Non-Extremal  Pure D3-Solution and Its 
 Regular and  Singular  Cases   }

In this section we trace the singular horizon
problem  of the special solution found in the previous section
 to a similar problem in the $P=0$ case, i.e.  to singularity 
of  certain non-extremal
generalizations of the regular  extremal 
D3-brane solution.

In general, the system of 
second-order differential equations  for $y$ \yuw\ and $z$ 
\zzy\ 
has an 
extra free  integration constant  --  an  
extra  parameter of  non-BPS deformation in addition to 
the constant $a$ in  $x$ in \xex.
The standard non-extremal D3 with regular horizon
\HS\  is a special 
case of a more general class of solutions.
One usually discards such more general solutions
  by imposing the condition that 
{\it the horizon should be regular}.  That condition 
is satisfied only for a special choice of two free
integration constants.

{}From the effective ``7-d black hole"  
point of view,  these more general solutions
correspond to the case when  an extra scalar 
(the radius  of the internal 3-torus) 
has  non-vanishing  asymptotic charge.
However,  regular black holes 
should not have scalar hair -- otherwise we get a 
singular horizon.
It is only when this extra scalar 
 charge is tuned to zero  that  we get  
a regular  non-extremal  D3-brane solution.

To consider  the non-extremal 
pure D3-brane case let us start with 
the general system of equations \yyy--\coln\
with $P=0$ and $f=0$.\foot{We could keep  $f$ non-constant  for $P=0$.
That  would introduce  an extra potential term 
in the  equations for the remaining fields.
Like the $\P\sim u$ case discussed in Appendix A, this  corresponds  to having 
an extra scalar charge and most likely leads
 to a singular solution.}
 To match the standard 
D3-brane solution we  shall also set $\P=0$
and $w=0$.\foot{The discussion of this section applies both to 
  $M_5= T^{1,1}$ and $M_5=S^5$.} 
Then  we are left with \xex\ and 
  the following  system:
\eqn\coi{ 
 y''  - 4 e^{8y} =0 \ , \ \ \ \ \
z''   - { 1 \ov 4} Q^2 e^{8z} =0  \ , }
i.e.  
\eqn\most{ x'=a \ , \ \ \ \ \ 
y'^2 = b^2  + e^{8y} \ , \ \ \ \ \ 
z'^2 = c^2   + q^2 e^{8z} \ , \ \ \ \  \ \  q\equiv  {1\ov 4} Q \ , 
}
with the  integration constants  $a,b,c$ 
related  by the zero-energy constraint \coln\ 
 \eqn\coon{
5b^2 - 3 a^2 - 2 c^2 =0   \ . 
}
Assuming that $a,b,c\geq 0$ (so that 
$y \to \infty$ for $u\to 0$) and that 
 $h,g$ in \iop\ approach  1    as $u\to 0$, 
 we find
\eqn\bau{    e^{4y} =  { b\ov \sinh 4b u } \ , \ \ \ \ \ 
 e^{4z} =  { c\ov  q\  \sinh 4 c (u + k)   } \ ,\ \ \ \ \ \ 
e^{4x} = e^{4au} \ , \ \ \ 
}
where $k$ is defined by 
\eqn\kiu{
 e^{ 4c k} = q^{-1}
 ( \sqrt{ q^2 + c^2} + c ) \equiv \g \ .  
}
Then (see  \iop) 
\eqn\kyti{
\r^4= e^{4y+4x} =  { 2b e^{4(a-b)u} \ov 1 - e^{-8bu} } \ , \ \ \ 
\ \ \ \ \    g= e^{-8au} \ ,  } 
\eqn\kyt{ h= e^{-4z- 4x} = 
e^{4(c-a)u}[ 1  + { q \ov 2c \g  }
 ( 1 - e^{- 8c u} ) ] 
    \ . 
}
At small $u$ (large $\r$) we have 
\eqn\yoi{ 
g= 1 - { 2a \ov \r^4} + ... \ ,  \ \ \ \ 
h = 1  + { \sqrt{ q^2 + c^2}  -a  \ov \r^4} + ...  
\ , \ \ \ \    \r^4 = { 1\ov 4 u} + ... \ . }

\subsec{\bf   Standard regular non-extremal D3-brane solution}
The 
{\it standard}  non-extremal  D3-brane solution 
\HS\ corresponds to the  case  when 
\eqn\specc{
b=c=a  \  , } 
i.e.~to  a line in the 2-parameter $(b,c)$ space.
Then the constraint \coon\ is satisfied, and 
 \bau, \kiu\ become 
\eqn\aau{    e^{4y} =  { a\ov \sinh 4a u  } \ , \ \ \ \ \ \
 e^{4z} =  { a\ov  q \sinh 4 a (u + k)  } \ ,\ \ \ \ \ \ \
e^{4x} = e^{4au} \ ,  
}
\eqn\iu{
\g=  e^{ 4a k} = q^{-1}
 ( \sqrt{ q^2 + a^2} + a ) \ . 
}
Note that  near the horizon ($u\to \infty$) 
\eqn\eess{
y= y_* - a u +  y_1  e^{-8au} + O( e^{-16au}) \ , \ \ \ \ \
z= z_*  - a u +  z_1  e^{-8au} + O( e^{-16au}) \ ,
}
\eqn\eee{
 y_* = { 1 \ov 4} \ln 2 a \ , \ \ \ \ 
y_1 =  { 1 \ov 4}\ , \ \ \ \ 
z_* = { 1 \ov 4} \ln {2 a\ov q \g} \ , \ \ \
z_1 =  { 1 \ov 4\g^2} \ ,  }
 while at large distances   ($u\to 0$)
\eqn\sss{
y= - { 1 \ov 4} \ln 4 u  - { 2 \ov 3} a^2 u^2 + O(u^3)  
\ , \ \ \ \ \
z= -   \bar q u  + O( u^2)  \ ,\ \ \ \ 
\bar q=  q\g^{-1} + a= \sqrt{q^2 + a^2} \ . 
}
It is { only}  for the choice  $b=c=a$ that  
  the metric \fop\ takes the 
standard non-extremal D3-brane form \refs{\HS,\dup} 
\eqn\mer{
ds^2_{10E} =  h^{-1/2} ( g dX_0^2 + dX_i dX_i)
+  h^{1/2}  [ g^{-1}  d\r^2 
+ \r^2 (d M_5)^2] \ , \ \ \ \ 
 }
\eqn\hji{  g= e^{-8x} = 1 - {2 a\ov  \r^{4}}  \  , 
 \ \ \ \ \ \ \ 
  \r^4= { 2 a \ov 1- e^{-8a u} } \ , }
\eqn\qqq{
h=  e^{-4z- 4x}= 1 + { \q \ov 
 \r^{4} }  \ ,  \ \
 \ \ \ \ \ \ \ \
 \q = \g^{-1} q  
=  \sqrt{ q^2 +  a^2 }  - a   \  .  }
In  an   often-used  parametrization 
$q= 2a \sinh \a \cosh \a$ and $\q= 2 a \sinh^2 \a$,
where the  charge $q$  is fixed in the extremal $a\to 0$ limit.
The mass  (density) of the solution is   
$M=  \sqrt{ q^2 +  a^2 }  + {3 \ov 4}  a\  >  \ q$,
so this is the standard  non-extremal black-brane 
 type solution with a {\it regular}  horizon.

Let us note also that the choice of $k=0$ in \aau\
leads to the  black hole in AdS  solution, 
where  we remove the asymptotically flat region. 
Then  the metric is \mer\  with 
\eqn\then{
  g= e^{-8x} = 1 - {2 a\ov  \r^{4}}  \  , 
 \ \ \ \ \ \ \ 
  \r^4= { 2 a \ov 1- e^{-8a u} } \ ,
\ \ \ \ \  h=  e^{-4z- 4x}= { q \ov 
 \r^{4} }  \ . }

\subsec{\bf   Special singular non-extremal D3-brane  solution}

The general solution  with arbitrary $b$ and $c$ 
reduces to the standard  extremal D3-brane background
only if we set $b$  and $c$ proportional to $a$, 
satisfying  the constraint \coon.

The simplest  special case  is $c=0$   where $z$ 
satisfies the 1-st order equation $z' = - q e^{4z}$
(cf. \most). Then from \bau\ we get 
\eqn\beu{    e^{4y} =  { b\ov \sinh 4b u  } \ , \ \ \ \ \
 e^{-4z} = 1 + 4 q u   \ ,\ \ \ \ \ \
e^{4x} = e^{4au} \ ,  \ \ \ \  \ \ b=\sqrt {3 \ov 5}  a\ , 
}
and thus 
 \eqn\tio{
\r^4= e^{4y+4x} =  { 2b e^{4(a-b)u} \ov 1 - e^{-8bu} } \ , \ \ \ \  
 h= e^{-4z- 4x} =  ( 1  + { 4 q  }  u  ) e^{-4au}
 \ , 
\ \ \  \   g= e^{-8au} \ . 
}
This solution, 
which is the {\it $P=0$  limit}  of the  solution \Buch\ 
derived in section 4,  
has a {\it singular} horizon at $u=\infty$. 

For small $u$  (large  distances) 
and in the limit $a\to 0$, we still get the
standard asymptotic extremal D3-brane behavior
\eqn\sta{ \r^4 = { 1\ov 4 u} + ... \ ,  \ \ \ \ \ 
g= 1 - { 2a \ov \r^4} + ... \ , \ \ \  
h = 1  +   { q-a \ov \r^4} + ...  
\ . }
If we define the horizon as the place where $G_{00}$ vanishes, then
its location is at infinite $u$. Note that there
$g\to 0$, but also $h \to 0$;  still
$ G_{00}= h^{-1/2} g \to 0$.

Since  $ b < a$,  the radial coordinate $\r$ is not 
a monotonic function of $u$:
it grows to infinity at both $u=0$ and $u=\infty$, 
having a minimum at finite $u$.
Therefore, $\r$ is not a good  (one-to-one) 
coordinate, and we must instead use 
$u$  to cover the entire space-time.
The same remark applies to the  fractional brane generalization 
of \tio, i.e. to the  solution of \Buch\  found
in  section 4.

It is only in the  case $b=c=a$ 
of the {\it standard} D3-brane solution \aau\ 
 that $\r$ is monotonic with 
 $\r(0) =\infty$, and $\r^4(\infty) = 2 a$ is the position
of the horizon.

The explicit form of   the  corresponding 
metric \fop\  in the special  case of  \tio\  is 
 \eqn\mio{
ds^2_{10E} = ( 1  + { 4 q  }  u  )^{-1/2} 
         ( e^{-6au}  dX_0^2 +  e^{2au} dX_i dX_i)
+    ( 1  + { 4 q  }  u  )^{1/2}   ds^2_6 \ , 
}
\eqn\lpo{ ds^2_6=   
({  b \ov  \sinh 4 b u})^{5/2}   
   du^2  
+   ({ b \ov  \sinh 4 b u})^{1/2}   (dM_5)^2 \ .} 
  This metric has a  horizon
  as well as   curvature
 singularity at $u=\infty$. (One can check  that 
$R_{mnkl} R^{mnkl}$ is divergent there.)

\newsec{\bf   Discussion  }

To summarize, 
the condition of regular horizon usually  
imposed on  black holes  excludes  the special non-extremal
D3-brane  solution discussed  in section 5.2.
This solution is special in that $z$ satisfies 
the same  1-st order equation (without an 
extra integration constant) as in the extremal case.
For that reason  it is this solution  that 
has the immediate simple fractional D3-brane 
generalization found 
in \Buch\ and 
described explicitly in section 4.
This solution 
has a singular horizon, which is  now not surprising since 
for  zero fractional D3-brane
charge $P$, it 
reduces {\it not} to the standard regular D3-brane solution of section 5.1
but to the  special singular solution of section 5.2.

Let us make the comparison with \Buch\ more explicit.
Just as for the $P=0$ case discussed in section 5.2, as we increase $u$ 
the radial coordinate $\rho$ reaches a minimum value $\rho_*$
and then starts
increasing again. In fact, $\rho(u)$ is exactly the same as in the
$P=0$ case, eq. \tio. In \Buch\ $\rho_*$ was incorrectly identified
as the horizon. But  it is not a horizon because $g=e^{-8x}$
is finite there -- the minimum value $\rho_*$ 
is an artefact of the coordinate choice.
In this solution the coordinate $u$ covers both branches of 
the $\rho$ coordinate. The singular horizon is located at $u=\infty$ where
$\rho=\infty$.

The above discussion shows 
that  in order for the  non-extremal generalization of the 
KT solution  to reduce to the standard 
black D3-brane in the $P\to 0$ limit,
the large $u$  asymptotics have to be (see  \aau, \eess)
\eqn\boun{
 \ u\gg 1 : \ \  \ \ \ 
x = a u \ , \ \ \ \   \ \ \ y\to - a u  + y_* \  , \ \  
\ \  \  
z\to - au +  z_*\ 
 ,    }
\eqn\bam{   w\to w_* \ , \ \  \ \ \ \   
\P\to \P_* \ ,  \ \  \ \ \ \ f\to  f_* \ . }
These asymptotics guarantee the existence of a regular
Schwarzschild horizon at $u=\infty$, and 
 it is natural to expect 
that $w,f$ and $\P$  
 have stationary points at this $u \to \infty$ 
 horizon.
Then it is easy to see that  our system of equations 
\yyy--\fef\ {\it and} the constraint 
\coln\  are indeed satisfied at large $u$. 
It is also not hard to check that turning
on $P$ makes a small perturbation on these asymptotics.

Since in the $P\rightarrow 0$ limit we  need to keep 
$c=a$ in the equation  \most\ for $z$,   
we are forced to 
give up the 1-st order equations for $z$ and $f$ \zfe, 
i.e.~the self-duality of the 3-forms.
In turn, the equations \ppp\ and  \yuw\ for $\P$ and $w$ 
then receive sources
and it is no longer possible  to have solutions with 
$\P=0$ and $w=0$. 
The reason for this special role of $w$ is that 
our ansatz  for the forms \har--\fiff\ breaks the symmetry between 
$\psi$ and other directions of $M_5$.
The lack of self-duality of the 3-forms
causes the $T^{1,1}$ to become squashed at finite $u$.

Now we face a formidable task of solving the complete
2-nd order system of equations  \yyy--\coln, 
looking for special non-singular solutions which 
interpolate between the KT asymptotics \kot,\ktt\ at small $u$ 
and the Schwarzschild horizon asymptotics  
\boun, \bam\ for large $u$.
Such solutions will be discussed in a future publication.

\bigskip
\noindent
{\bf Acknowledgements}
\bigskip
We are grateful to Steve Gubser and Peter Ouyang for
useful discussions. This research 
was supported in part by the NSF under Grant No. PHY99-07949.
The work of A.~B. was  also supported in part 
by the NSF under Grant No. PHY97-22022. 
The work of C.~P.~H. and I.~R.~K. was also supported in part by the NSF
under Grant No. PHY98-02484.
L.P.Z. would like to acknowledge the Office of the Provost at the University of
Michigan and DOE for support.
The work of A.T. is supported in part by
the DOE grant  DOE-FG02-91R-40690, PPARC SPG grant,
and INTAS project 991590.

\appendix{A}{\bf Non-BPS solution  with  non-constant dilaton  and $w=0$}

Here we discuss singular solutions with a non-constant
 dilaton, 
\eqn\diol{
\Phi = 4 p u \ . }
 Now the zero-energy constraint is
 \eqn\tko{ 5 y'^2     - 5 w'^2   
 -    e^{8y} ( 6 e^{-2w} - e^{-12 w} ) = 3 a^2 + 2 p^2  
\ . }
We find
\eqn\sool{
e^{4y} = { b \ov \sinh 4bu} \ , \ \ \ \ \  \ \ 5b^2 = 3 a^2 + 2p^2\
.   }
The functions $f$ and $z$ are then determined from \zfe.

If we keep $a$ and $p$ independent  and 
 take $P=0$, 
then we do not get
the standard non-extremal  D3-brane 
 solution but rather its   generalization
with  non-zero dilatonic charge.  If $a=0$ but 
the non-vanishing dilaton charge $p\not=0$, 
we recover 
the conifold analog of the 
singular  generalization of the extremal D3-brane 
solution discussed in \refs{\SK,\Gub}.
 The fractional 3-brane case with non-zero $P$ 
still leads to a singular solution.

It is therefore  necessary 
 to relate $p$ and $a$, i.e. to 
 demand that 
\eqn\choi{ p= m a \ , \ \ \ \ \   b= n a\ , \ \ \ \ 
n=[ {1\ov 5} (3 +2  m^2 )]^{1/2} \ .  }
The relation \choi\  is needed in order to  get back 
the KT solution (and not its other non-BPS generalization
with running dilaton)
in the limit  $a=0$.
Then $a$ plays the role of the non-extremality parameter 
which ``drives'' the solution away from 
the BPS point of the  KT  background \ktt.

One particularly simple  possibility 
 is  $m=1$, i.e. 
\eqn\cho{ p=b=a\ .  }
Then eqs. \zfe\ become
\eqn\fff{
f'  + {aP e^{4au} \ov \sinh 4au}  =0 \ ,
\ \ \ \ \ \ \ \ 
(e^{-4z})'  =  Q + 2 Pf \ ,  }
so that 
\eqn\fee{
f= f_*  - { P \ov 4} \ln ( e^{8a u} -1 ) \ ,  }
which of course  reduces to the KT expression in \kot\ 
in the 
BPS $a=0$ limit ($f_0 = f_*  - { P \ov 4} \ln 8a )$.
  $f$  goes  in the same 
$\log u$  way for small $u$ (large distances) 
and linearly with $u$ for large $u$ (small distances), 
a novel behaviour.

The  equation for $z$ gives 
\eqn\zer{
e^{-4z} = C +  K_*  u  -  2 a P^2 u^2 
-   {P^2 \ov 16 a} {\rm Li}_2(e^{-8au}) \ , }
$$ C=1 +{ P^2 \pi^2 \ov 96 a} \ , \ \ \ \ \ \ \
K_*= Q + 2f_* P \ , $$ 
where $ C$ is such that $z(0)=0$
to have  the standard 
long distance limit.
Explicitly, for small $u$ we reproduce  the
KT asymptotic 
behaviour
\eqn\kett{
e^{-4 z} = 1   - 
 { P^2  \ov 2 } u \log u  + 
[K_* + \ha P^2 (1 - \ln 8a)] u + 
O(u^2)  \ , }  
while for large $u$
\eqn\zeri{
e^{-4z} = -  2 a P^2 u^2  + K_* u  + O(e^{-8au})
 \ , }
indicating the presence of a  special point 
at finite $u$ where  $e^{-4z} $  vanishes.\foot{Numerical  analysis  confirms that 
there exists a finite value of $u$ were 
 $ z$ goes to infinity.
The  behaviour of $e^{-4z}$  is  similar to that of the  
$ 1 + u - u^2$ function: its starts at 1, grows  reaching a  
maximum, and then goes to zero at  
approximately 
$u_s=1.617$.}

This point  is a  curvature singularity.
According to \fop, \iop,  $h = e^{-4z-4x}$, 
so that  $h=0$ at {\it finite}
 $u$. Note that $g=e^{-8x}$  and $\r= e^{y+x}$  are 
 still finite there, so this is  a   {\it naked   singularity}.
Just as in the KT case,   the derivative of $z$  or 
$(e^{-4z})'= K(u)$  becomes zero  at $u=u_0$ 
 before we reach 
that singular point: 
$ e^{8 a u_0} = 1 + 2 P^{-1} e^{K_0}$. 
Since the derivative  of $e^{-4z}$ changes sign, that means 
$e^{-4z}$ reaches a maximum at $u=u_0$ and then goes to zero.

Above we considered the case of $p > 0$ when 
the string coupling $e^\P = e^{4pu}$  grows  at small distances.
One finds a somewhat nicer behaviour of the metric 
in the opposite case of  $m=-1$, i.e.
\eqn\chii{
p=-b=-a \ . }
Now for large $u$ 
the  string coupling becomes weak  and
$f\rightarrow  f_*$.
This produces a solution with a singular horizon at
$u=\infty$.  For example, if we choose $Q+2P f_*=0$, then
$e^{-4z}$ approaches a constant for large $u$.
The horizon is singular because
the string-frame  metric becomes 
$$
ds_{string}^2 \rightarrow 
e^{-8au} dX_0^2 + dX_i dX_i + e^{-12 a u} du^2 + e^{-4au} (dM_5)^2
\ , 
$$
so that the 
 longitudinal volume stays finite but the transverse 
volume vanishes
at the horizon.

\vfill\eject
\listrefs
\end